# UNIFYING IMAGE QUALITY ASSESSMENT DATASETS: MOSAIQ-500K AND MOSAIQ-BENCH

*Wenbo Yang*, Zhongling Wang+, Jialu Xu*, Jinghan Zhou*, Zhou Wang**

*University of Waterloo, + AI Center-Toronto, Samsung Electronics

**ABSTRACT**

Image quality assessment (IQA) datasets use different subjective protocols and rating scales, so their scores are not directly comparable. The lack of a common perceptual scale hinders multi-dataset training and precludes direct inter-dataset evaluation. We address this by conducting a new subjective experiment and using its ratings as perceptual anchors to fit monotonic mappings that place the existing scores of 23 IQA datasets on a common quality scale while preserving within-dataset rankings. The resulting dataset, MOSAIQ-500K, contains over 500,000 images and is, to our knowledge, the largest IQA dataset with perceptually aligned subjective scores. We also propose MOSAIQ-Bench, an inter-dataset benchmark, and use it to evaluate 31 IQA methods, revealing substantial gaps between intra- and inter-dataset performance, particularly on authentic distortions. The aligned scores offer a key advantage: replacing specialized multi-dataset training mechanisms with standard regression losses yields comparable intra-dataset accuracy and better inter-dataset performance. MOSAIQ thus provides a practical foundation for combining heterogeneous IQA datasets to train more generalizable IQA models. Code and dataset will be made public upon acceptance.

## 1. INTRODUCTION

Image quality assessment (IQA) relies on datasets annotated with subjective quality scores (SQSs) for model training and evaluation. Collectively, existing datasets cover diverse image content, distortion types, and quality levels, offering the potential to train more generalizable models and evaluate them under broader conditions. However, they differ in subjective test protocols, rating scales, and image distributions, so their scores are not directly comparable. Even after simple normalization, equal scores from different datasets need not indicate equal perceptual quality. This misalignment hinders combining existing datasets, both for inter-dataset evaluation and for multi-dataset training.

Existing IQA evaluation computes correlations between predicted quality scores (PQSs) and SQSs separately for each dataset, even in cross-dataset settings where models are tested on unseen datasets. We refer to this as *intra-dataset* evaluation. In contrast, *inter-dataset evaluation* computes correlations over images pooled from multiple datasets, testing whether PQSs are consistent across datasets with different content, distortion, and quality distributions, which intra-dataset evaluation cannot reveal (Fig. 1). However, inter-dataset evaluation requires SQSs on a unified quality scale, which existing datasets do not provide.

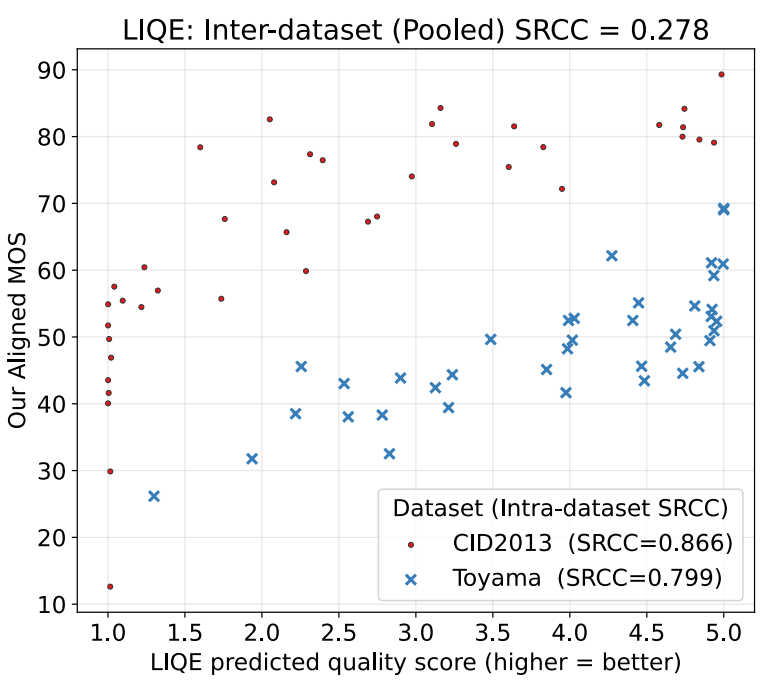

**Fig. 1**. LIQE [1]'s predictions achieve relatively high SRCC on both CID2013 and Toyama without being trained on either, which is considered a good performance in the traditional cross-dataset sense. Yet direct inter-dataset evaluation using MOSAIQ reveals that it consistently scores CID2013 images lower than Toyama images of equal subjective quality.

The lack of a common quality scale also complicates multi-dataset training. Hence, most learning-based IQA models are trained on a single dataset, while some methods circumvent the misalignment with specialized training mechanisms. Compare2Score [2] and DeQA-Score [3] convert SQSs into coarse categorical labels, sacrificing fine-grained score information, whereas UNIQUE [4], LIQE [1], VisualQuality-R1 [5], and ReLIQS [6] learn from intra-dataset pairwise comparisons using Thurstone's model. MonotonicIQA [7] takes a more direct approach, learning a separate mapping from PQSs to each dataset's MOSs. These workarounds provide no explicit supervision on how images from different datasets compare perceptually. Consequently, current multi-dataset training approaches do not guarantee consistent predictions across datasets.

We sample 1,039 images spanning the score ranges of 26 IQA datasets for a subjective experiment. After screening, 36,760 ratings from 40 subjects on 919 images from 23 datasets are retained. The images span diverse content, distortions, and quality levels. Since their ratings are collected directly, inter-dataset evaluation on the combined set is not affected by errors in the score alignment described next. Using these MOSs, we propose the MOS-Aligned IQA Benchmark (**MOSAIQ-Bench**), and use it to evaluate 31 IQA methods, revealing substantial discrepancy between intra- and inter-dataset performance, particularly for no-reference (NR) methods (Section 3.1). While prior works [8, 9] considered dataset alignment,

+Work done at the University of Waterloo.

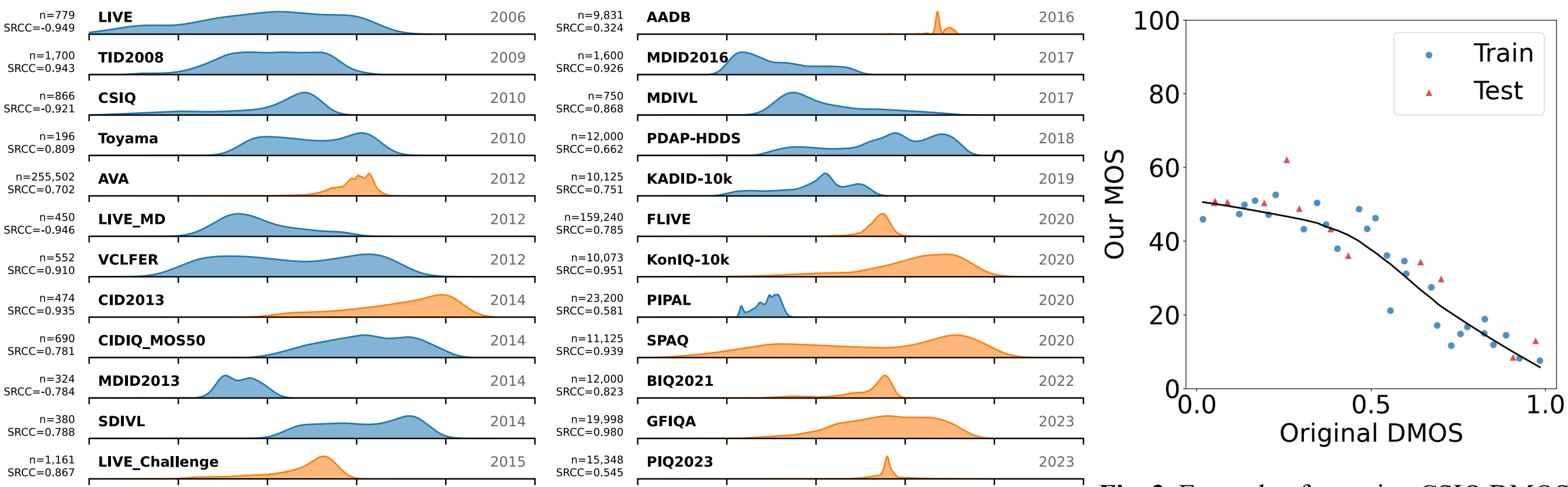

**Fig. 3**. Example of mapping CSIQ DMOS values onto the common MOS scale.

**Fig. 2**. Distribution of quality scores from each dataset (mapped to our common scale). The dataset size and the SRCC between the original quality scores and our MOSs are also shown.

we are the first to distinguish between intra- and inter-dataset evaluation.

Using the collected ratings as perceptual anchors, we fit monotonic mappings that place each dataset's original scores on a common scale while preserving within-dataset quality rankings. We call scores on the common scale unified quality scores (UQSs). Applying these mappings to all images in the 23 datasets yields the MOS-Aligned Image Quality dataset (**MOSAIQ-500K**), to our knowledge, the largest IQA dataset with perceptually aligned subjective scores (over 500,000 images). MOSAIQ-500K is much larger than what are produced prior alignment [8, 9, 10], and it simplifies and enhances multi-dataset training: UQSs provide direct supervision on quality comparisons across datasets, even for IQA models without specialized multi-dataset training mechanisms for handling misaligned scales. In our experiments, models with and without such mechanisms can be trained on MOSAIQ-500K, achieving competitive intra-dataset performance and improved inter-dataset performance on unseen datasets (Section 3.2). In summary, our contributions are:

- **MOSAIQ-Bench**: an inter-dataset benchmark showing that intra-dataset performance does not predict inter-dataset performance.
- **MOSAIQ-500K**: an IQA dataset constructed by realigning 23 existing datasets onto a common quality scale.
- Evidence that training on **MOSAIQ-500K** allows specialized multi-dataset mechanisms to be replaced with a plain regression loss and while improving inter-dataset performance.

## 2. CONSTRUCTION AND ANALYSIS OF MOSAIQ-500K

FR-IQA methods can only be benchmarked on datasets that pair each distorted image with a pristine reference, whereas NR-IQA methods can be evaluated on datasets containing only distorted images. We refer to these two types of datasets as FR datasets (FRDSs) and NR datasets (NRDSs), respectively. We collected 15 FRDSs [10, 11, 12, 13, 14, 15, 16, 17, 18, 19, 20, 21, 22, 23, 24] and 11 NRDSs [25, 26, 27, 28, 29, 30, 31, 32, 33, 34, 35], all named in Fig. 2. From each dataset, we sampled 40 distorted images (39 for Toyama), with their original SQSs approximately uniformly distributed over the dataset's SQS range.

### 2.1. Subjective test

We conduct a subjective test on the sampled images from all datasets with 42 subjects. Each subject rates the quality of each image in one session, without seeing reference images, on a 0–100 scale, using a workstation with a BenQ PD2706U 27-inch 4K monitor. Each monitor is factory-calibrated to $\Delta E \leq 0.13$ and operates in sRGB mode at full brightness. To compensate for the 4K monitor's higher DPI, an image is displayed at 2x its original size unless it is too large to fit the screen, in which case it is scaled to fit. We recruited 23 male and 19 female subjects from the University of Waterloo, 25 of whom reported experience in image processing, computer vision, or photography. All subjects passed a vision screening confirming normal or corrected-to-normal vision.

Each session begins with an eight-image training sequence, during which a researcher explains the rating scale and discusses each image with the subject, focusing on technical quality. Next, a 22-image stabilizing sequence helps subjects calibrate their rating scale, before the main test images are shown in random order. Subjects are not informed of the stabilizing sequence. Both training and stabilizing images are sampled from the unified datasets but excluded from the main test. Six test images are randomly repeated two or three times to facilitate subject screening.

### 2.2. Screening and analysis of subjects and datasets

Let $\mathcal{I}_r$ denote the set of indices of repeated images, and $\mathcal{X}_{si}$ the set of scores given by subject $s$ to image $i$. The wide-rating count of subject $s$ is defined as

$$w_s = |\{i \in \mathcal{I}_r : \max(\mathcal{X}_{si}) - \min(\mathcal{X}_{si}) > 1.25\sigma_s\}|, \quad (1)$$

where $\sigma_s$ is the standard deviation of all ratings given by subject $s$. A subject with $w_s > 2$ is deemed unreliable and excluded from the analysis. This removes two male subjects, leaving 40. BT.500 screening [36] is not applied, as it is recommended only for tests with fewer than 20 subjects.

For each remaining subject, the average rating for each repeated image is used as the final rating for that image. Unlike [37], we do not remove outlier ratings, since deviations from the mean often reflect individual preferences rather than random noise [38], as discussed in Section 2.5. The ratings of all remaining subjects are then averaged to obtain the MOS of each image.

**Table 1**. Comparison of the selected score-realignment method with alternative methods. Due to space limits, the coefficient of determination ($R^2$) between the mapped SQSs and our MOSs on the test portions is only shown on some datasets. The mean is computed across all datasets.

| Method | AVA | CID | CSIQ | KADID | KonIQ | LIVE | CLIVE | LIVE-MD | SPAQ | TID08 | TID13 | Mean |
|---|---|---|---|---|---|---|---|---|---|---|---|---|
| Proposed | .731 | .766 | .848 | .317 | .919 | .823 | .721 | .893 | .924 | .866 | .449 | **.751** |
| Logistic 4 | .732 | .758 | .832 | .320 | .920 | .820 | .734 | .885 | .927 | .856 | .323 | .737 |
| Logistic 5 | .730 | .758 | .829 | .284 | .918 | .821 | .736 | .903 | .927 | .880 | .327 | .737 |
| CDF Matching | .783 | .792 | .806 | .138 | .934 | .784 | .717 | .901 | .922 | .826 | .375 | .725 |

After the subjective test, we compute the SRCC between our MOSs and the original SQSs of each dataset (shown in Fig. 2). PIPAL [19] and PIQ2023 [33] derive their SQSs from pairwise (2AFC) comparisons [8], which do not guarantee that the final scores are comparable across different scenes. AADB [25] targets aesthetic rather than technical quality. We therefore exclude these three datasets from the subsequent tests. Although AVA [26] is also labeled "aesthetic", its scores correlate highly with technical quality, so we retain it. The remaining 23 datasets form MOSAIQ-500K, and their 919 rated images form MOSAIQ-Bench.

### 2.3. IQA dataset score realignment

**Problem Statement** Each IQA dataset provides SQSs as MOSs, DMOSs or in other forms, and it is commonly assumed that if they are positively correlated with human perception, they can be realigned onto a common scale. Let $S_{di}$ denote our MOS for image $i$ in dataset $d$, and $R_{di}$ its original SQS. Since the purpose is to align existing SQSs to *unified quality scores* (UQSs) rather than to train another IQA model, the mapping function $f_d$ for each dataset $d$ takes only $R_{di}$ as input, so that $f_d(R_{di}) \approx S_{di}$. Because our goal is to place scores on a common scale without reordering images by quality within a dataset, $f_d$ must be monotonic. Furthermore, to prevent distinct $R_{di}$ values from collapsing to the same mapped score, the monotonicity must be strict for the most of the domain of $f_d$.

**Train-Test Split** For each NRDS, images with and without our MOSs are separately split into training and test sets at 3:1 ratio, with five-bin stratified sampling for the former. For each FRDS, the 3:1 split is based on the reference images, so that all distorted versions of a reference image are assigned to the same partition. Datasets whose reference images are derived from the overlapping sets of source images (LIVE [10], TID2008 [21], TID2013 [22], MDID2013 [15], and Toyama [23]) are split at the source-image level, and a split is rejected if its test partition does not contain $10 \pm 2$ images with our MOSs.

**Regression** To enforce the desired properties of $f_d$, we apply LOWESS followed by isotonic regression and PCHIP interpolation [39]. The resulting UQSs $f_d(R_{di})$ are used as the labels in MOSAIQ-500. An example mapping is shown in Fig. 3.

**Findings** The mapping results (Fig. 2) confirm a common belief in the IQA research community: different IQA datasets cover vastly different ranges of image quality, indicating the necessity of realignment. For example, almost all images in LIVE [10] have lower quality than those in UHD-IQA [35]. Additionally, some IQA datasets have concentrated SQSs, such as FLIVE [29] and PIQ [33], while others have a more uniform distribution, such as LIVE [10] and SPAQ [34]. Additionally, the SQSs of some datasets, such as FLIVE [29] and PIQ [33], are concentrated, whereas those of others, such as LIVE [10] and SPAQ [34], are more uniformly distributed.

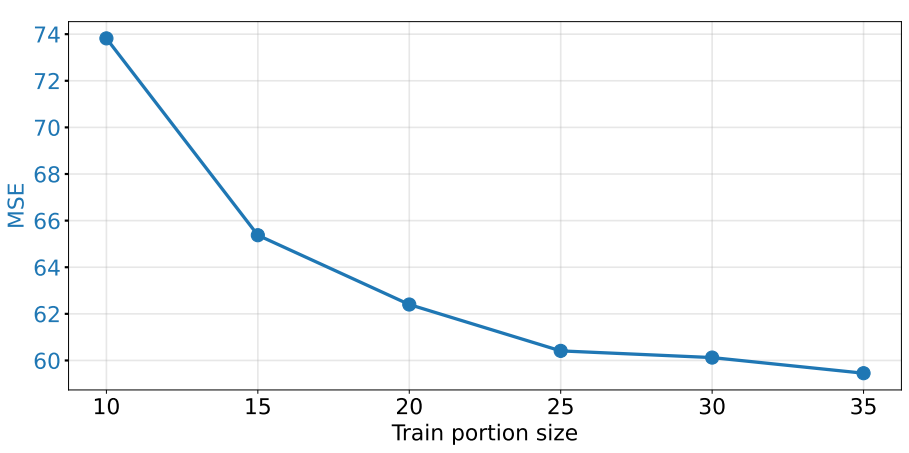

**Fig. 4**. MSE between the mapped SQSs and our MOSs on the test portion of each split.

### 2.4. Discussion: alternative fitting methods

It is trivial to derive that all strictly monotonic mapping functions with the correct direction yield the same SRCC between $f_d(R_{di})$ and $S_{di}$. Therefore, we only consider methods that gives an $f_d$ that is strictly monotonic on the majority of the range of $R_{di}$. The quality of these $f_d$ is best measured by the coefficient of determination ($R^2$) between $f_d(R_{di})$ and $S_{di}$. Table 1 compares the selected method with several alternatives. The selected method achieves the highest average $R^2$. Another question is whether 30 training examples per dataset are sufficient for the regression. We repeat the regression with 10, 15, 20, 25, 30, and 35 training examples per dataset and compute the MSE between the mapped SQSs and our MOSs in the test portion of each split. The procedure is repeated 100 times and the average is reported in Fig. 4, which shows that the curve plateaus at at 25-35, indicating that 30 training examples are sufficient.

### 2.5. Discussion: individual variation of opinion scores

We use SigLIP 2 (NaFlex) So400m/16 [40] to extract an image feature $\boldsymbol{v}_i$ for each image. Following [38], we model each subject's rating $S_{si}$ as the MOS plus a residual $r_{si} = \boldsymbol{z}_i'\boldsymbol{\theta}_s$ and random noise:

$$S_{si} = S_i + r_{si} + \varepsilon_{is}, \tag{2}$$

where $\boldsymbol{z}_i = [\boldsymbol{v}_i; S_i, 1]$ is the image characterization, and $\boldsymbol{\theta}_s$ is the subject characterization. We estimate $\boldsymbol{\theta}_s$ by solving a ridge-regression problem:

$$\boldsymbol{\theta}_s = \arg\min_{\boldsymbol{\theta}_s} \sum_i \|Z\boldsymbol{\theta}_s - \boldsymbol{r}_s\|_2^2 + \lambda\boldsymbol{\theta}_s' R\boldsymbol{\theta}_s, \tag{3}$$

where $Z = [\boldsymbol{z}_1; \boldsymbol{z}_2; ...; \boldsymbol{z}_N]$ and $R = \mathrm{diag}([1, ..., 1, 0])$. We also fit two baselines: (a) $\boldsymbol{z}_i$ set to a constant vector, and (b) each subject's residuals shuffled across images. The coefficient of determination of residual prediction is 0.405, moderately higher than the baselines' (0.331 and 0.314), indicating that individual variation in opinion scores can be partially explained by subjects' preferences for image features, in addition to simple bias and random noise. Hence, removing individual rating from reliable subjects discards useful information and should be avoided.

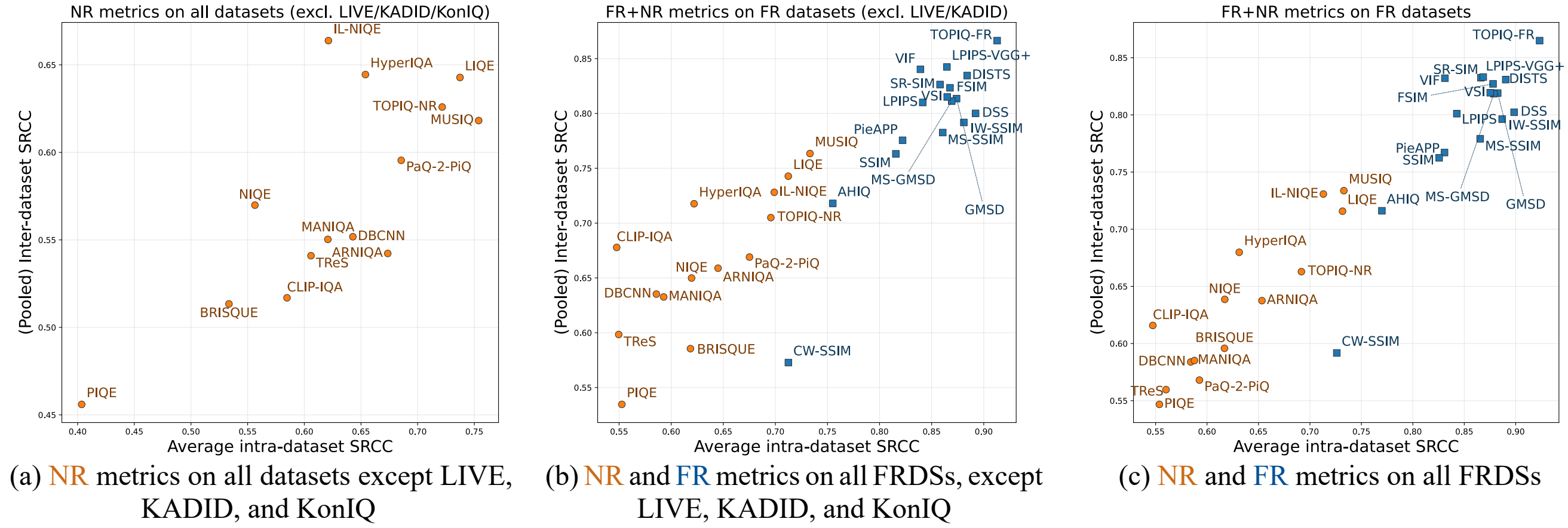

(a) NR metrics on all datasets except LIVE, KADID, and KonIQ (b) NR and FR metrics on all FRDSs, except LIVE, KADID, and KonIQ (c) NR and FR metrics on all FRDSs

**Fig. 5**. MOSAIQ-Bench results for existing IQA methods. The x-axis shows the intra-dataset SRCC (the mean of the per-dataset SRCCs over the included datasets), and the y-axis shows the inter-dataset SRCC, computed against our MOSs over the pooled images.

## 3. INTER-DATASET PERFORMANCE BENCHMARK AND ENHANCEMENTS FOR IQA METHODS

### 3.1. MOSAIQ-Bench: inter-dataset performance benchmark

Our MOSs enable direct evaluation of the inter-dataset performance of existing IQA methods. Fig. 5 shows results for 14 NR-IQA and 17 FR-IQA methods, using default implementations and weights from piq [41] and PyIQA [42]. All benchmarked models are training-free, trained without SQSs, trained on datasets outside MOSAIQ, or trained on LIVE [10], KADID [13], or KonIQ [31]. Hence, the three sets are excluded from the tests in Fig. 5(a-b). For completeness, Fig. 5(c) reports results on all datasets, including the three sets.

Three observations emerge from the benchmark. First, except for a few outliers, inter- and intra-dataset performance are similar on FRDSs, possibly due to FRDSs' similar distortion simulation. CW-SSIM is a notable exception: its substantially lower inter-dataset performance suggests limited practicality. Second, NR methods' inter- and intra-dataset performance have larger discrepancy on NRDSs than on FRDSs, indicating that NRDSs differ substantially in distortion distribution and that NR methods vary in robustness to these differences. Third, IL-NIQE [43] is an interesting outlier: despite relatively low intra-dataset performance on NRDSs, it achieves the best inter-dataset performance among NR methods, on par with some FR methods.

### 3.2. Improving inter-dataset performance

UQSs in MOSAIQ give models not designed for multi-dataset training a direct route to it, and render the dedicated mechanisms of models that already support it unnecessary. We evaluate three representative methods: LIQE [1], MonotonicIQA [7], and HyperIQA [44]. The first two use different multi-dataset training strategies with fine-grained labels, while HyperIQA was not designed for it. Since LIQE requires distortion labels its authors did not provide for other datasets in MOSAIQ, we train it only on LIVE, CSIQ, KADID, KonIQ, and CLIVE. For better comparison, both other methods are trained on the same five sets. Table 2 shows three retrained variants of each method. We report intra-dataset (mean per-dataset SRCC) and inter-dataset (SRCC over pooled images) performance, separately for the five seen datasets and the unseen ones.

**Table 2**. Inter-dataset (and intra-dataset) SRCC of IQA methods trained with and without our UQSs. On the five seen sets, models are tested on the test portion only.

| Model | Variant | Seen 5 sets | Unseen sets |
|---|---|---|---|
| **Models designed for multi-dataset training** | | | |
| MonotonicIQA | V1: Orig. Method | .7415 (.9278) | .7836 (.7681) |
| | V2: V1 + our UQSs | .7719 (.9279) | .8217 (.7824) |
| | V3: V2 - Map. mod. | **.7823** (.9126) | **.8591** (.7841) |
| LIQE | V1: Orig. Method | .7262 (.9437) | .7262 (.7944) |
| | V2: V1 + our UQSs | **.7878** (.9363) | .7885 (.7576) |
| | V3: V2 - Pair loss | .7655 (.9383) | **.7912** (.7899) |
| **Model not designed for multi-dataset training** | | | |
| HyperIQA | V1: Orig. Method | .5561 (.6722) | .5394 (.6311) |
| | V2: Naive norm. | .6710 (.8915) | .5372 (.6149) |
| | V3: our UQSs | **.7088** (.9058) | **.5707** (.6391) |

For MonotonicIQA and LIQE, V1 is the original method trained on the five datasets, V2 replaces the original SQSs with our UQSs (which enables inter-dataset comparisons in LIQE's loss), and V3 replaces their multi-dataset training strategy with a simple L1/L2 loss. Specifically, MonotonicIQA-V3 drops the per-dataset network mapping PQSs to original SQSs, and LIQE-V3 replaces the pairwise comparison loss with L2. Table 2 indicates that our UQSs improve inter-dataset performance for both, although they already attempted to address score misalignment. V3′s similar intra-dataset and superior inter-dataset performance relative to V1 shows that the alignment makes the original multi-dataset strategies unnecessary.

For HyperIQA, V1 is the original model retrained on KonIQ, V2 is trained on the five unified datasets with naive per-dataset score normalization, and V3 is trained using our aligned SQSs. V3 performs best in both intra- and inter-dataset terms on seen and unseen datasets.

## 4. CONCLUSION

We presented a systematic method for unifying IQA datasets with inconsistent subjective score scales. Using ratings collected on a common subjective test for 1,039 images sampled from 26 datasets, we created a new benchmark **MOSAIQ-Bench**. MOSAIQ-Bench shows that strong intra-dataset performance does not necessarily translate to strong inter-dataset performance, particularly for NR-IQA models on authentic distortions. Furthermore, we estimated monotonic mappings that align scores across datasets, yielding **MOSAIQ-500K**, a unified IQA dataset. Training on MOSAIQ-500K further improves inter-dataset performance on unseen datasets. Our work provides a practical foundation for combining heterogeneous IQA datasets, evaluating IQA models' robustness, and training more generalizable IQA models. The code and dataset will be available at `ivc.uwaterloo.ca/projects/unifying_iqa_datasets/` upon acceptance.

## 5. ACKNOWLEDGEMENTS

This study has received clearance from the University of Waterloo's Human Research Ethics Board (#30196) and is supported in part by NSERC Discovery Grant (RGPIN-2024-05154) and Canada Research Chair Program (CRC-2025-00266). LLMs (incl. Grok, Claude, Kimi, ChatGPT, and DeepSeek) were used to accelerate the development of some parts of the codebase and to edit the text. AI-generated code has been properly tested.

## 6. REFERENCES

[1] W. Zhang, G. Zhai, Y. Wei, X. Yang, and K. Ma, "Blind Image Quality Assessment via Vision-Language Correspondence: A Multitask Learning Perspective," in *CVPR*, 2023.
[2] H. Zhu *et al.*, "Adaptive Image Quality Assessment via Teaching Large Multimodal Model to Compare," in *NeurIPS*, Curran Associates, Inc., 2024. doi: 10.52202/079017-1024.
[3] Z. You, X. Cai, J. Gu, T. Xue, and C. Dong, "Teaching Large Language Models to Regress Accurate Image Quality Scores Using Score Distribution," in *CVPR*, 2025.
[4] W. Zhang, K. Ma, G. Zhai, and X. Yang, "Learning to blindly assess image quality in the laboratory and wild," in *ICIP*, 2020.
[5] T. Wu, J. Zou, J. Liang, L. Zhang, and K. Ma, "Visualquality-r1: Reasoning-induced image quality assessment via reinforcement learning to rank," *NeurIPS*, 2026.
[6] H. E. Gedik, S. Gupta, and A. Bovik, "Learning Where to Look and How to Judge: Resolution-agnostic Image Quality Assessment with Quality-aware Saliency." [Online]. Available: https://arxiv.org/abs/2608.01730
[7] Z. Feng, K. Zhang, S. Jia, B. Chen, and S. Wang, "Learning from mixed datasets: A monotonic image quality assessment model," *Electronics Letters*, vol. 59, no. 3, p. e12698, 2023.
[8] M. Pérez-Ortiz, A. Mikhailiuk, E. Zerman, V. Hulusic, G. Valenzise, and R. K. Mantiuk, "From Pairwise Comparisons and Rating to a Unified Quality Scale," *IEEE TIP*, 2020.
[9] A. Mikhailiuk, M. Pérez-Ortiz, D. Yue, W. Suen, and R. K. Mantiuk, "Consolidated Dataset and Metrics for High-Dynamic-Range Image Quality," *IEEE Transactions on Multimedia*, vol. 24, pp. 2125–2138, 2022.
[10] H. Sheikh, M. Sabir, and A. Bovik, "A Statistical Evaluation of Recent Full Reference Image Quality Assessment Algorithms," *IEEE TIP*, vol. 15, no. 11, pp. 3440–3451, 2006.
[11] X. Liu, M. Pedersen, and J. Y. Hardeberg, "CID: IQ–a new image quality database," in *ICISP*, 2014, pp. 193–202.
[12] E. C. Larson and D. M. Chandler, "Most apparent distortion: full-reference image quality assessment and the role of strategy," *Journal of Electronic Imaging*, 2010.
[13] H. Lin, V. Hosu, and D. Saupe, "KADID-10k: A Large-scale Artificially Distorted IQA Database," in *QoMEX*, 2019.
[14] D. Jayaraman, A. Mittal, A. K. Moorthy, and A. C. Bovik, "Objective quality assessment of multiply distorted images," in *ACSSC*, 2012.
[15] K. Gu, G. Zhai, X. Yang, and W. Zhang, "Hybrid no-reference quality metric for singly and multiply distorted images," *IEEE Transactions on Broadcasting*, 2014.
[16] W. Sun, F. Zhou, and Q. Liao, "MDID: A multiply distorted image database for image quality assessment," *Pattern Recognition*, doi: https://doi.org/10.1016/j.patcog.2016.07.033.
[17] S. Corchs and F. Gasparini, "A multidistortion database for image quality," in *CCIW*, 2017, pp. 95–104.
[18] T.-J. Liu, H.-H. Liu, S.-C. Pei, and K.-H. Liu, "A High-Definition Diversity-Scene Database for Image Quality Assessment," *IEEE Access*, 2018, doi: 10.1109/ACCESS.2018.2864514.
[19] G. Jinjin, C. Haoming, C. Haoyu, Y. Xiaoxing, J. S. Ren, and D. Chao, "PIPAL: a large-scale image quality assessment dataset for perceptual image restoration," in *ECCV*, 2020.
[20] S. Corchs, F. Gasparini, and R. Schettini, "No Reference Image Quality classification for JPEG-Distorted Images," *Digital Signal Processing*, doi: 10.1016/j.dsp.2014.04.003.
[21] N. Ponomarenko, V. Lukin, A. Zelensky, K. Egiazarian, M. Carli, and F. Battisti, "TID2008-a database for evaluation of full-reference visual quality assessment metrics," *Advances of modern radioelectronics*, vol. 10, no. 4, pp. 30–45, 2009.
[22] N. Ponomarenko *et al.*, "Image database TID2013: Peculiarities, results and perspectives," *Signal processing: Image communication*, vol. 30, pp. 57–77, 2015.
[23] Y. Horita, K. Shibata, and Y. Kawayoke, "MICT Image Quality Evaluation Database." Accessed: Aug. 27, 2015. [Online]. Available: https://mict.eng.u-toyama.ac.jp/mictdb.html
[24] A. Zarić *et al.*, "VCL@ FER image quality assessment database," *automatika*, vol. 53, no. 4, pp. 344–354, 2012.
[25] S. Kong, X. Shen, Z. Lin, R. Mech, and C. Fowlkes, "Photo aesthetics ranking network with attributes and content adaptation," in *ECCV*, 2016, pp. 662–679.
[26] N. Murray, L. Marchesotti, and F. Perronnin, "AVA: A large-scale database for aesthetic visual analysis," in *CVPR*, 2012, pp. 2408–2415. doi: 10.1109/CVPR.2012.6247954.
[27] N. Ahmed and S. Asif, "BIQ2021: A large-scale blind image quality assessment database," *Journal of Electronic Imaging*, vol. 31, no. 5, p. 53010, 2022.
[28] T. Virtanen, M. Nuutinen, M. Vaahteranoksa, P. Oittinen, and J. Häkkinen, "CID2013: A database for evaluating no-reference image quality assessment algorithms," *IEEE TIP*, 2014.
[29] H. Niu, "FLIVE Database." Accessed: Sep. 09, 2026. [Online]. Available: https://github.com/niu-haoran/FLIVE_Database/tree/master
[30] S. Su *et al.*, "Going the Extra Mile in Face Image Quality Assessment: A Novel Database and Model," *IEEE TMM*, 2023.
[31] V. Hosu, H. Lin, T. Sziranyi, and D. Saupe, "KonIQ-10k: An Ecologically Valid Database for Deep Learning of Blind Image Quality Assessment," *IEEE TIP*, vol. 29, pp. 4041–4056, 2020.
[32] D. Ghadiyaram and A. C. Bovik, "Massive online crowdsourced study of subjective and objective picture quality," *IEEE TIP*, vol. 25, no. 1, pp. 372–387, 2015.
[33] N. Chahine, S. Calarasanu, D. Garcia-Civiero, T. Cayla, S. Ferradans, and J. Ponce, "An image quality assessment dataset for portraits," in *CVPR*, 2023.
[34] Y. Fang, H. Zhu, Y. Zeng, K. Ma, and Z. Wang, "Perceptual quality assessment of smartphone photography," in *CVPR*, 2020.
[35] V. Hosu, L. Agnolucci, O. Wiedemann, D. Iso, and D. Saupe, "UHD-IQA Benchmark Database: Pushing the Boundaries of Blind Photo Quality Assessment." [Online]. Available: https://arxiv.org/abs/2406.17472
[36] ITU-R, "Methodologies for the Subjective Assessment of the Quality of Television Images," Recommendation BT.500–15, May 2023.
[37] K. Ma *et al.*, "Group maximum differentiation competition: Model comparison with few samples," *IEEE TPAMI*, 2018.
[38] J. Zhou and Z. Wang, "GC360IQ: Generic-to-Individualized Quality Assessment for Stitched 360-Degree Panoramas." [Online]. Available: https://arxiv.org/abs/2609.08118
[39] F. N. Fritsch and J. Butland, "A Method for Constructing Local Monotone Piecewise Cubic Interpolants," *SIAM Journal on Scientific and Statistical Computing*, 1984.
[40] M. Tschannen *et al.*, "Siglip 2: Multilingual vision-language encoders with improved semantic understanding, localization, and dense features," *arXiv preprint arXiv:2502.14786*, 2025.
[41] S. Kastryulin, J. Zakirov, D. Prokopenko, and D. V. Dylov, "PyTorch Image Quality: Metrics for Image Quality Assessment." [Online]. Available: https://arxiv.org/abs/2208.14818
[42] C. Chen and J. Mo, "IQA-PyTorch: PyTorch Toolbox for Image Quality Assessment." Accessed: Jan. 01, 2026. [Online]. Available: https://github.com/chaofengc/IQA-PyTorch
[43] L. Zhang, L. Zhang, and A. C. Bovik, "A feature-enriched completely blind image quality evaluator," *IEEE TIP*, vol. 24, no. 8, pp. 2579–2591, 2015.
[44] S. Su *et al.*, "Blindly Assess Image Quality in the Wild Guided by a Self-Adaptive Hyper Network," in *CVPR*, Jun. 2020.